\documentclass[natbib]{svmult}

\usepackage{makeidx}
\usepackage{graphicx}
\usepackage{multicol}

\newcommand{\Paramesh}{{\sc{Paramesh}}~}
\newcommand{\FLASH}{{\sc{Flash}}~}

\makeindex

\title*{Efficiency Gains from Time Refinement on 
       AMR Meshes and Explicit Timestepping}
\titlerunning{Efficiency Gains from Time Refinement}

\author{L.~J.~Dursi\inst{1} \and 
M.~Zingale\inst{2}}

\institute{Dept.\ of Astronomy \& Astrophysics, The University of Chicago, Chicago, IL  60637 ({\tt ljdursi@flash.uchicago.edu}) \and 
Dept.\ of Astronomy \& Astrophysics, The University of California, Santa Cruz, Santa Cruz, CA 95064 ({\tt zingale@ucolick.org})}

\begin{document}
\maketitle

\begin{abstract}
Block-structured AMR meshes are often used in astrophysical fluid
simulations, where the geometry of the domain is simple.
We consider potential efficiency gains for time sub-cycling, or
time refinement (TR), on Berger-Collela and oct-tree AMR meshes for
explicit or local physics (such as explict hydrodynamics), where the
work per block is roughly constant with level of refinement.   We note
that there are generally many more fine zones than there are
coarse zones.  We then quantify the natural result that any overall
efficiency gains from reducing the amount of work on the relatively few
coarse zones must necessarily be fairly small.  Potential efficiency
benefits from TR on these meshes are seen to be quite limited except
in the case of refining a small number of points on a large mesh ---
in this case, the benefit can be made arbitrarily large, albeit at the
expense of spatial refinement efficiency.  
\end{abstract}


\section{Introduction}

\subsection{Block-Structured AMR}
Adaptive mesh refinement on rectangular grids (henceforth AMR) was
introduced in \cite{bergeroliger}, and improved for conservation laws in
\cite{bergercollela}, henceforth BC89.  In the patch-based meshes of the
sort described in BC89, the patches increase in resolution by a fixed even
integer factor $N$.  One can place a finer patch anywhere in the domain
of a `parent' patch of one fewer level of refinement.  A patch is not
required to have only a single parent, but must be completely contained
within patches of the next lowest level of refinement.  Note that these
meshes are non-conforming; the face of a zone in a parent patch will
abut $N$ faces in the child patch.  A final restriction in the nesting
of the meshes is that there must be at least one zone of the next lower
level refinement about the perimeter of a patch.

Another mesh we will consider here is an oct-tree mesh (quad-tree in
2-d, binary tree in 1d), such as is implemented in the \Paramesh package
\cite{paramesh} used in the \FLASH code \cite{flashcode}.  This oct-tree
mesh is a more restrictive version of an $N=2$ patch-based mesh as
described in BC89.  If a block needs additional resolution somewhere
in its domain, the entire block is halved in each coordinate direction,
creating $2^d$ children, where $d$ is the dimensionality.  Leaf blocks
are defined to be those blocks with no children, and are thus at the
bottom of the tree --- they are the finest-resolved blocks in their
region of the domain.  Frequently, only leaf blocks are evolved to
compute the solution to the equations, since a refined parent block's
domain is completely spanned by its children.

The only difference between the two meshing approaches of immediate
interest is the resulting different refinement patterns.  We will use
`patch' and `block' interchangeably in this paper.

\subsection{Time Refinement}

\begin{figure}[hH]
\begin{center}
\includegraphics[width=.9\textwidth]{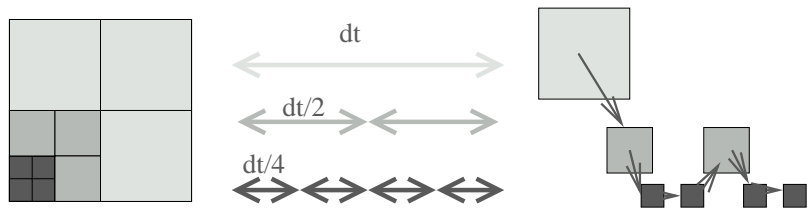}
\end{center}
\caption{A structured AMR mesh containing blocks at three
         different levels of refinement, showing the order of operations
         (far right) of an explicit time evolution algorithm.  The largest
         block is evolved at the system timestep, and smaller blocks are
         subcycled at smaller timesteps.  Between evolution at different
         levels of the mesh, time averaging and flux corrections must
         be done --- these are not shown here.}
\label{fig:tmrwcurve}
\end{figure}

In BC89, the timestep set by the data on
the finest mesh is used to evolve that data, and data on the coarser
meshes is evolved at a multiple thereof so that there is a constant ratio at each
level $l$ of $\Delta t_l$ to $\Delta x_l$.   The assumption here is that there is
one roughly spatially constant characteristic speed throughout the entire domain,
so that the maximum allowable timestep at any given resolution is
directly proportional to the size of the mesh for any given block or
patch.   When coupled with
the assumption in structured AMR of some fixed jump in
refinement between levels, this makes for a very natural time evolution
algorithm, shown pictured in Figure~\ref{fig:tmrwcurve} for a mesh with
three different levels of refinement, with resolution jumps by constant
factors of $N$; shown is $N=2$.

Here the largest blocks are evolved at some system timestep $dt$,
and smaller blocks are `subcycled' at proportionally smaller timesteps.
This defines a `work function' for each block; the finest blocks must be
evolved every sub-timestep so we take their work value to be 1 times the
number of zones in the block or patch;  the blocks one level of refinement
`up' need only be evolved every $N$ sub-timesteps, so that their work
value is $1/N$ times the number of zones, etc.  The work function for
an entire mesh is the sum of the work values of each block or patch in
the mesh.

There are costs associated with this time refinement (hereafter TR).
Memory is needed to store information at multiple timesteps.
There are overheads from extra copies and time-centering of fluxes.
The modified time-structure of work leads to load-balance issues in
parallel jobs.  Further complicating parallel performance is increased
communication complexity (although, it is to be pointed out, not
necessarily increased communication).

Nonetheless, one might hope that these costs are outweighed by the time
savings of not evolving large blocks at unnecessarily small timesteps;
in the example of Figure~\ref{fig:tmrwcurve}, of evolving the larger
blocks at timesteps of $dt$ or $dt/2$ instead of $dt/4$.  
As a first step to quantify the possible benefits, we
estimate the reduction in computational cost in simple cases
\S\ref{sec:analysis}.  We then use the same approach to examine
meshes from simulations performed with a
tree-based mesh in \S{\ref{sec:data}}.  In
our final section we summarize our results.

\section{Simple Mesh Configurations}
\label{sec:analysis}

\newcommand{\nblocks}{N_{\mathrm{blocks}}}
\newcommand{\Wtmr}{W_{\mathrm{TR}}}
\newcommand{\Wnontmr}{W_{\mathrm{noTR}}}

Here we calculate both the number of evolved blocks in a simple mesh, and a
weighted sum representing the ideal amount of work done by a TR method,
using the work function described in the previous section.
We then calculate a work ratio, $R$ --- the amount of work that
would be done by the idealized TR divided by that done with no
time refinement.  With no time refinement, each block must be
stepped through each sub-timestep, so that the amount of work done
is simply the number of blocks; thus, the work ratio is simply
(TR~work~function)/(number~of~blocks).  For $R = 1$, there is
no reduction in work; for $R < 1$, TR reduces the amount of computational work.

One can interpret the work ratios as performance metrics for the TR,
assuming that -- all physics benifits from the
time subcycling in proportion to the reduction in number of blocks evolved
each step; the memory overhead from TR is unimportant; all larger blocks
actually {\emph{can}} be evolved at timesteps of larger size in proportion
to their physical size; there is no single-processor overhead from TR
from memory copies or flux averaging; there is no parallel overhead from
increased complexity in communications; and there is no parallel from
increased load-balancing issues.

\subsection{Point refinement}
\label{subsec:pointrefine}

\begin{figure}[ht]
\begin{center}
\includegraphics[width=.2\textwidth]{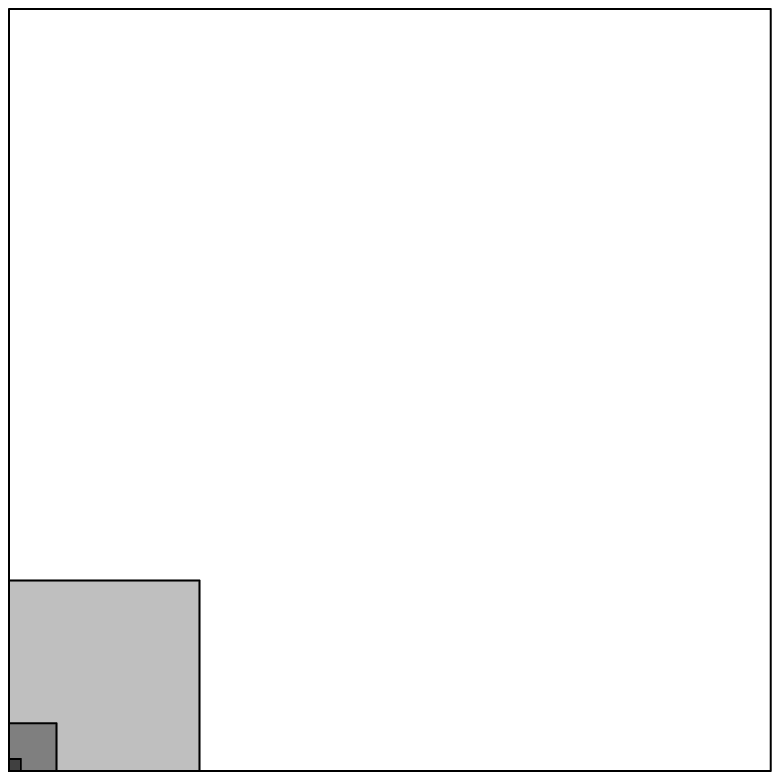}
\includegraphics[width=.2\textwidth]{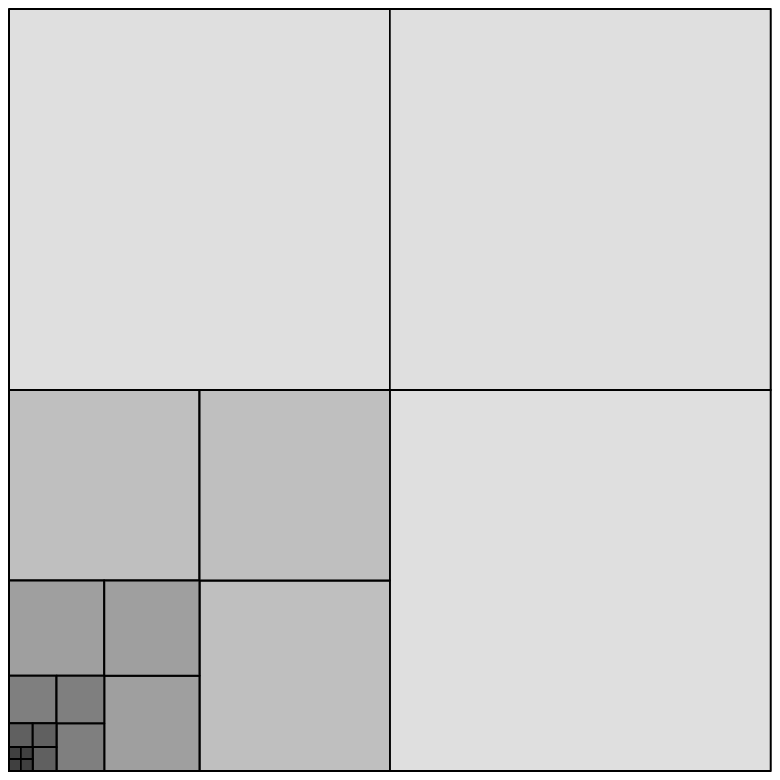}
\hskip .3 in
\includegraphics[width=.2\textwidth]{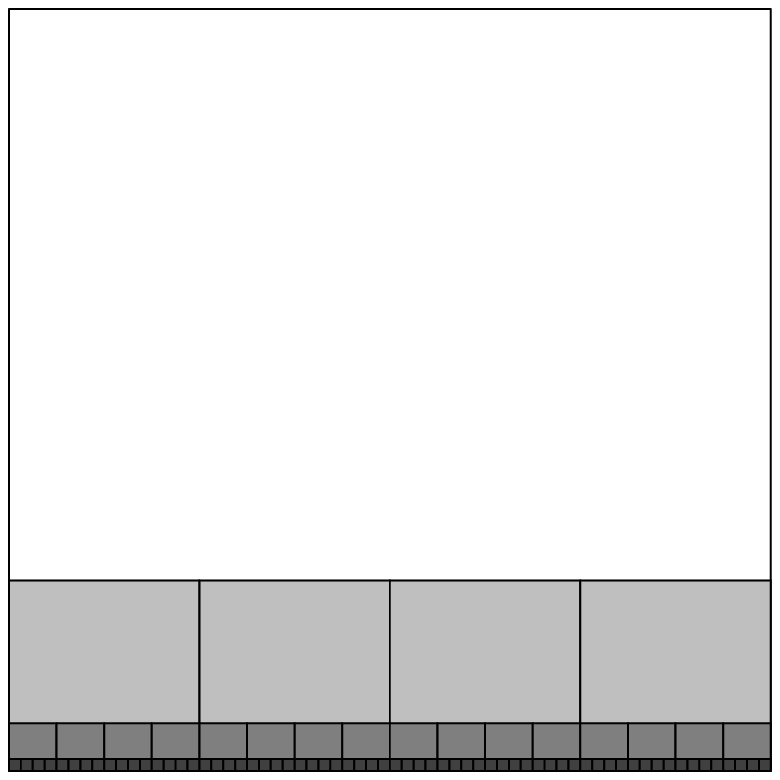}
\includegraphics[width=.2\textwidth]{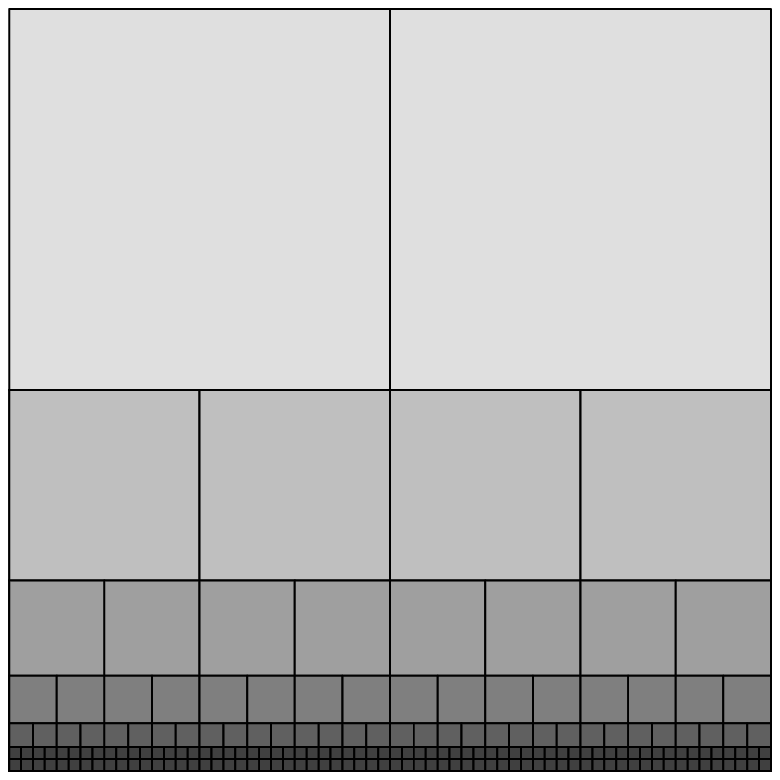}
\end{center}
\caption{Fully refining a zero-thickness point
         with an idealized patch-based type mesh (far left) and an oct-tree
         mesh (left);  Fully refining an interface with a patch-based mesh
         (right) and oct-tree mesh (far right).  For the patch-based mesh, it is assumed that
         a patch can be placed anywhere on existing patches, with some
         fixed integral increase in resolution (shown here is $N=4$, $L=3$).
         For the oct-tree mesh, $N$ is fixed at 2, and shown is $L = 5$.}
\label{fig:meshes-point} 
\end{figure}

The best case for efficiency gains for spatial
refinement is clearly one isolated point of refinement.   For a
patched-based mesh, we imagine refinements as shown on the left of in
Figure~\ref{fig:meshes-point}.  

We begin with domains of length one in all directions.  The
completely unrefined domain is defined to be at level $l=1$ of
refinement.  Consider placing increasingly fine patches
at the corner, until we resolved the finest scale $\Delta x$
we wished.  If this requires $L-1$ more levels of refinement, each
decreasing the zone size by an integer factor $N$, then we have $\Delta
x \sim (1/N)^{L-1}$.  We will assume $\Delta x \ll 1/2$.

We consider the mesh in terms of the smallest uniform unit --- for the
oct-tree mesh, this is a single block, which will be of size $n_x \times
n_y \times n_z$ zones.  For the patch-based mesh, since the patches
can be of arbitrary size (and shape), we consider zones individually.
(Because we are not modelling guardcell filling, we can safely ignore
the fact that these zones are actually components of patches).  Thus, in
the results given below, an oct-tree mesh with (say) $8 \times 8$-zone
blocks at a maximum refinement $L = 5$ has the same resolution as a
patch-based mesh with $L = 8$.

The amount of work required by a non-TR code with only explicit or
local solves will, by assumption, be the same for each block, so that
$\Wnontmr = \nblocks$.  The amount of work with time
refinement, $\Wtmr$, will be a weighted sum of blocks.
For the pointwise-refined patch mesh, the number of blocks will simply
be $\nblocks = L$, as there is only one block per level.  The amount of
TR work is

\begin{equation}
\Wtmr  =  \sum_{l=1}^{L} 1 \cdot \left ( \frac{1}{N} \right )^{L-l}  \sim  \frac{N}{N-1}.
\end{equation}

Thus the work ratio will be
\begin{equation}
R = \frac{\Wtmr}{\Wnontmr} = \frac{\Wtmr}{\nblocks} =  \frac{N}{L(N-1)}.
\end{equation}

For ideal spatial AMR, where one can do all the refinement with
only one jump, $L = 2$, and so the amount of work done by a TR
algorithm is bounded from below at $1/2$ of the non-TR work.   At the
other limit, for a much less aggressive AMR with $N=2$, then the work
can be made an arbitrarily small fraction of the non-TR algorithm,
with $R = 2/L$ --- but note that this work ratio is achieved only by
operating on $L/2$ times as many blocks as in the best case for spatial
AMR.  

The oct-tree meshes refining
on a point is shown on the right of Figure~\ref{fig:meshes-point}.
In this case, there are $2^d$ highest refined blocks in the corner,
with the rest of the $2^d-1$ surrounding blocks at the next highest
refinement, surrounded by the $2^d-1$ surrounding blocks at the next
highest level of refinement, and so on.

Thus the total number of leaf blocks is 
\begin{equation}
\nblocks = (2^d) + \sum_{l=L-1}^{1} {\left ( 2^d-1 \right )} =  2^d (L - 1) - L + 2
\end{equation}

Weighting them by the amount of work,
\begin{equation}
\Wtmr = (2^d) + \sum_{l=L-1}^{1} {\left ( \frac{(2^d-1)}{2^{(L-l)}} \right )}  \sim  2^{(d+1)}-1
\end{equation}
making the work ratio
\begin{equation}
R  =  \left \{ \begin{array}{cl} 3/L &  1d \\
                                   7/(3L-2) & 2d \\
                                   15/(7L-6) & 3d 
                  \end{array}
        \right .
\end{equation}

As with the patch-based result, this ratio goes to zero for arbitrarily
large $L$.  These results are similar to the $N=2$ patch-based result, but
TR performs better here, and the spatial refinement worse ---  both  of
these are due to the fact that the oct-tree mesh generates more intermediate-level
blocks.

\subsection{Planar Interface Refinement}
\label{subsec:planerefine}

The refinement of an interface is shown on the right of 
Figure~\ref{fig:meshes-point}.  In the patch-based case, we continually
place a grid of $N$-by-1 (in 2d) or $N^2$-by-1 (in 3d) patches along
the interface, until the required resolution is achieved.  

In this case, performing the same calculation as in the previous section, one
obtains
\begin{equation}
R   \approx 1 - \frac{N-1}{N^d - 1} .
\end{equation}

Here, there is a fixed lower bound for the amount of work the TR can
achieve.  In the spatially-optimal large-$N$ limit, no work is saved
at all: $R \rightarrow 1$.   At the other limit, for $N=2$, in 2d,
$R \rightarrow 2/3$; in 3d, $R \rightarrow 6/7$.

In the oct-tree mesh we begin with one block at
the coarsest level.  It must be divided into 4 in this 2D
example, or, in general, $2^d$.   Half of these blocks will be 
further refined.  This continues until we reach the maximum
level of refinement.  The work ratio one finds is
\begin{equation}
R  =  \left \{ \begin{array}{cl} 7/9 & 2d \\ 45/49 & 3d \end{array} \right .
\end{equation}

In the point-refinement case of the previous subsection, a point of zero volume
needed to be refined; as a result, there were the same number of blocks at each
level, and thus a significant time savings could be obtained by doing less work
at the coarser blocks.  However, as we begin to see here, as soon as a non-trivial
volume of the mesh needs to be refined, there is significantly less savings to
be had.

\subsection{Circular Region Refinement}

\begin{figure}[ht]
\begin{center}
\includegraphics[width=.2\textwidth]{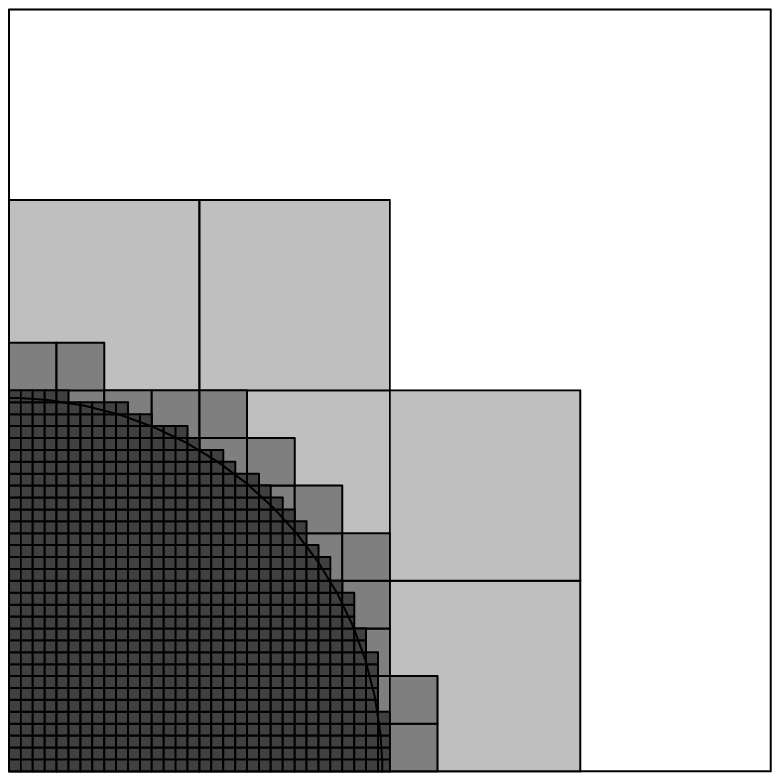}
\hskip .3 in
\includegraphics[width=.2\textwidth]{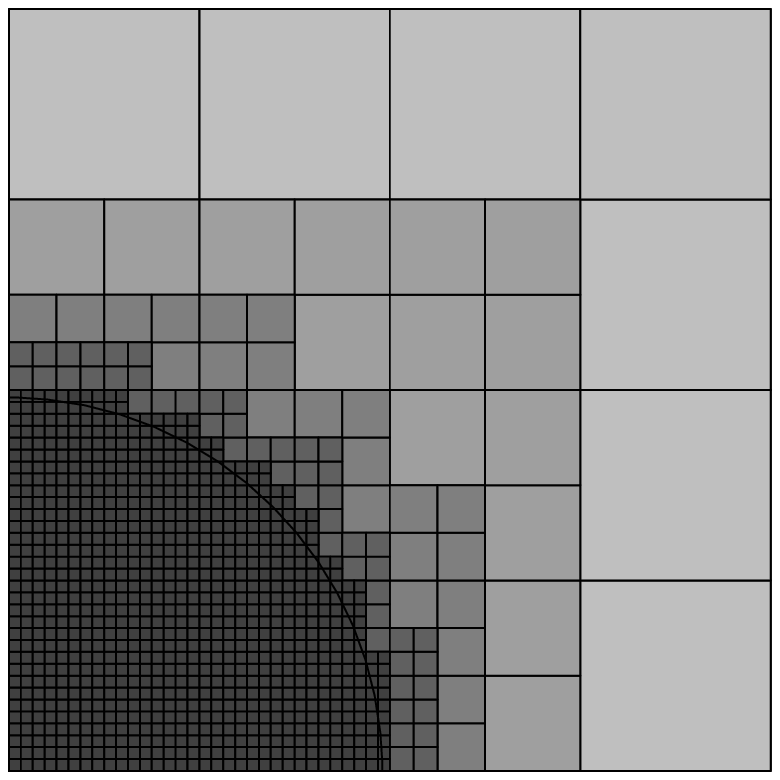}
\end{center}
\caption{Fully refining the interior of a circle, shown here with radius
         of $0.49$ of the
         box size, with an idealized patch-based type mesh (left) and
         an oct-tree mesh (right).   The patch-based mesh shown has $L
         = 3$ and $N = 4$.  For the oct-tree mesh, $N$ is fixed at 2,
         and shown is $L = 6$.}
\label{fig:meshes-curve-region}
\end{figure}

The loss of efficiency gains when a non-zero fraction of the mesh must
be refined is even clearer when a region, rather than an interface,
is fully refined.   In Fig.~\ref{fig:meshes-curve-region} we see
the results of fully refining the interior of a quarter-circle with the center
at one of the corners of the domain.  Clearly, the number of finest
blocks greatly outnumber intermediate or large blocks, so one might
guess that there is very little efficiency gain that can be had from
reducing work on the larger blocks.

\begin{table}[ht]
\begin{center}
\begin{tabular}{lrrrrrrrr}
{} & L=2 & 3 & 4 & 5 & 6 & 7 & 8 \\
\hline
r = 0.0 &  0.786 & 0.625 & 0.510 & 0.426 & 0.363 & 0.316 & 0.279 \\
    0.1 &  0.786 & 0.625 & 0.510 & 0.510 & 0.638 & 0.765 & 0.879 \\
    0.2 &  0.786 & 0.625 & 0.625 & 0.714 & 0.806 & 0.895 & 0.940 \\
    0.5 &  0.962 & 0.843 & 0.851 & 0.888 & 0.931 & 0.963 & 0.981 \\ 
    0.9 &  1.    & 0.973 & 0.962 & 0.962 & 0.973 & 0.982 & 0.989 
\end{tabular}
\end{center}
\caption{Work ratio for a 2d Oct-tree mesh with a circular region
of radius $r$ (in units of the domain) completely refined.}
\label{table:2d-oct-circregion}
\end{table}

Because in this case the refinement pattern is complicated enough that
the process must be iterated to check that each zones neighbors are
no further than one level of refinement appart, we do not provide
analytic work ratios.  Tables~\ref{table:2d-oct-circregion} and
\ref{table:2d-bo-circregion} show the work ratios for an Oct-Tree mesh
and an $N=2$ patch-based mesh in refining a circular region of radius $r$.
Again, the $r=0$ results reproduce the expected point refinement, but as
soon as a non-zero radius must be refined,  the efficiency gains drop
significantly further than in the case of only refining an interface,
as more small blocks are needed to refine a region than the interface.
In Table~\ref{table:2d-bo-circregion} we also show results for the patch
based mesh with $N=4$; we see as in previous sections that for the
same resolution, increasing $N$ (which increases the spatial efficiency
of AMR) decreases the possible gains from time subcycling.

\begin{table}[ht]
\begin{center}
\begin{tabular}{lrrrrrrr||rrr}
{} & N=2, L= 2 & 3 & 4 & 5 & 6 & 7 & 8 & N=4, L=2 & 3 & 4 \\
\hline
r = 0.0 &      0.583 & 0.468& 0.387& 0.328& 0.283& 0.249& 0.221 & 0.438 & 0.332 & 0.510 \\
    0.1 &      0.583 & 0.468& 0.444& 0.552& 0.658& 0.754& 0.802 & 0.719 & 0.891 & 0.510\\
    0.2 &      0.583 & 0.548& 0.618& 0.694& 0.768& 0.806& 0.833 & 0.812 & 0.914 & 0.625\\
    0.5 &      0.75  & 0.737& 0.763& 0.798& 0.825& 0.840& 0.848 & 0.896 & 0.938 & 0.851\\
    0.9 &      0.847 & 0.827& 0.826& 0.833& 0.842& 0.848& 0.852 & 0.938 & 0.947 & 0.962\\
\end{tabular}
\end{center}
\caption{Work ratio for a 2d patch-based mesh, $N=2$ and $N=4$, with a circular region
of radius $r$ (in units of the domain) completely refined.}
\label{table:2d-bo-circregion}
\end{table}

\section{Meshes from simulations}
\label{sec:data}

\newcommand{\ramr}{R_{\mathrm{AMR}}}

The calculations of the previous section are for very simple refinement
geometries.  In this section, we apply the same work function used in
\S\ref{sec:analysis} to the output of previous actual AMR simulations
which use oct-tree based meshes for AMR.   We continue to assume the
same idealized performance results of the previous section.

\begin{figure}[hHt]
\begin{center}
\includegraphics[width=.6 \textwidth]{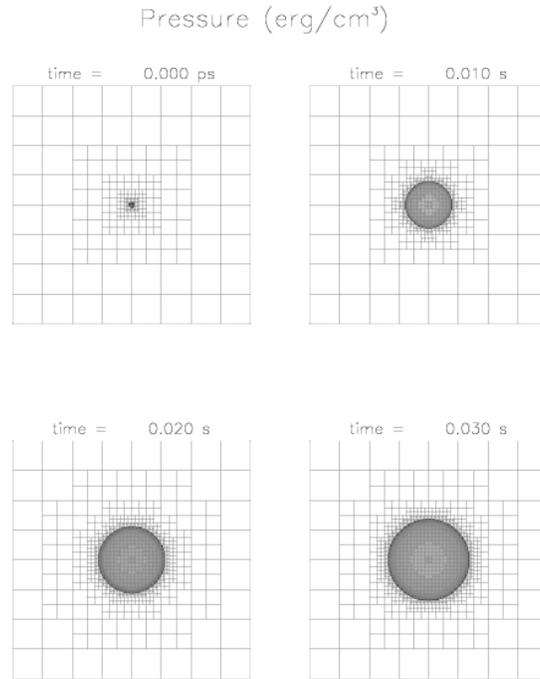}
\end{center}
\caption{The mesh of a Sedov explosion, from the \FLASH setup test described in
\cite{flashcode}, with a maximum of 8 levels of refinement.   Each
block shown contains $8 \times 8$ zones.}  
\label{fig:sedov}
\end{figure}

We begin with examining results from a standard test problem,
a Sedov explosion \cite{sedov}, as included with the \FLASH code and described
in \cite{flashcode}.  In this simulation, a high pressure at a point
causes a spherical shock wave to expand outwards; this is analogous
to the circular region analysis of the previous section.  The adaptive mesh for 
different stages of this simulation in 2d are shown in \ref{fig:sedov}.

Results from the meshes shown are tabulated in
Table~\ref{tab:sedovresults}.   The number of blocks listed in the table
is the number of `leaf' blocks -- {\emph {e.g.}}, the blocks that are
actually evolved.    Also given in the table is the work ratio ($R$)
and the work ratio of spatial AMR to a uniform mesh at the highest
resolution ($\ramr = \Wnontmr / W_{\mathrm{uniform}}$).  We include
$\ramr$ to compare the relative importance of performance gains for
the spatial refinement and the time subcycling.

TR provides a large performance gain initially, when there is only
one point that is refined.  However, consistant with previous results,
immediately as the point becomes a region of non-zero measure, idealized
performance gains drop to $30\%$--$10\%$.  Regardless
of the refinement, the TR provides a very small performance enhancement
compared to that of the spatial refinement.

\begin{table}[hHt]
\begin{center}
\begin{tabular}{c|rll}
time & $\nblocks$ & $R$ & $\ramr$ \\
\hline
0.00 & 256 & 0.426 & 0.0156 \\
0.01 & 892 & 0.805 & 0.0544 \\
0.02 & 1552 & 0.835 & 0.0947 \\
0.03 & 2092 & 0.874 & 0.127 \\
\end{tabular}
\end{center}
\caption{Results from simulations of a Sedov explosion.  Listed
at different evolution times are the number of leaf blocks in the mesh,
the work ratio, and the work ratio for spatial AMR to uniform grid.}
\label{tab:sedovresults}
\end{table}

The reason for the small predicted efficiency gains, consistent with
the discussion of the previous section, is that there quickly become
more fine blocks than coarse blocks in the simulation.   By the last
frame shown in Figure~\ref{fig:sedov}, there are no blocks being evolved
at the the coarsest level of refinement, and indeed 80\% of the blocks
are at the highest level of refinement.  Thus, even if all other blocks
required zero work to evolve, we could only achieve a $20\%$ speedup.

\begin{figure}[hHt]
\begin{center}
\includegraphics[angle=90,width=.7\textwidth]{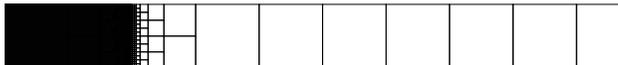}
\end{center}
\caption{Half of the domain for the initial condition of a detonation, where the
         long domain is refined nowhere except at a sharp interface.
         The domain originally consists of a top-level mesh of $1 \times
         20$ blocks.  This mesh is then refined at an interface.  Shown
         is the meshes 6, zoomed in near
         the interface.  Not shown are 10 coarsest blocks to the right.}
\label{fig:celldet} 
\end{figure}

Next we consider an interface problem -- a 2d detonation that will
eventually undergo a cellular instability.  These simulations are from
results published in \cite{celldet2d}.   A mesh is shown in 
Figure~\ref{fig:celldet}.  This
corresponds almost exactly to the idealized interface problem of the
previous section, but here the domain is very
long in one direction, increasing the number of low-cost coarsest
blocks in the domain.  This change in distribution of blocks means that
this problem can benefit more from TR.   The numerical results are
shown in Table~\ref{tab:celldet}.

\begin{table}[hHt]
\begin{center}
\begin{tabular}{c|rlll}
Max refinement & $\nblocks$  & $R$ & $\ramr$ \\
\hline
4 & 62 & 0.633 & 0.0484 \\
5 & 110 & 0.688 & 0.0215 \\
6 & 206 & 0.727 & 0.0101 \\
7 & 398 & 0.751 & 0.00486
\end{tabular}
\end{center}
\caption{Results from initial conditions for a 2-d detonation problem,
         as in Figure~\ref{fig:celldet}.  $R$ is less than
         the $7/9$ calculated in the previous section, because of the large number
         of extra coarsest  blocks added to the domain.}
\label{tab:celldet}
\end{table}

Here we see TR's efficiency gains actually decrease with increasing
resolution, and also see a familiar pattern of TRs efficiency gains
going in the opposite direction of spatial AMR efficiency gains. 
Even at the resolution where TRs efficiency gains are largest, they are
much smaller than the improvement from using spatial AMR.

\begin{figure}[hHt]
\begin{center}
\includegraphics[height=.625\textwidth,angle=90]{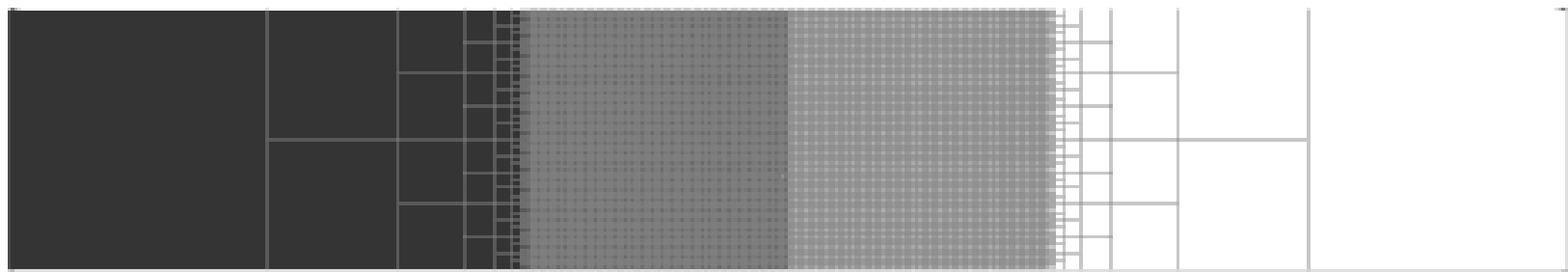}
\includegraphics[height=.625\textwidth,angle=90]{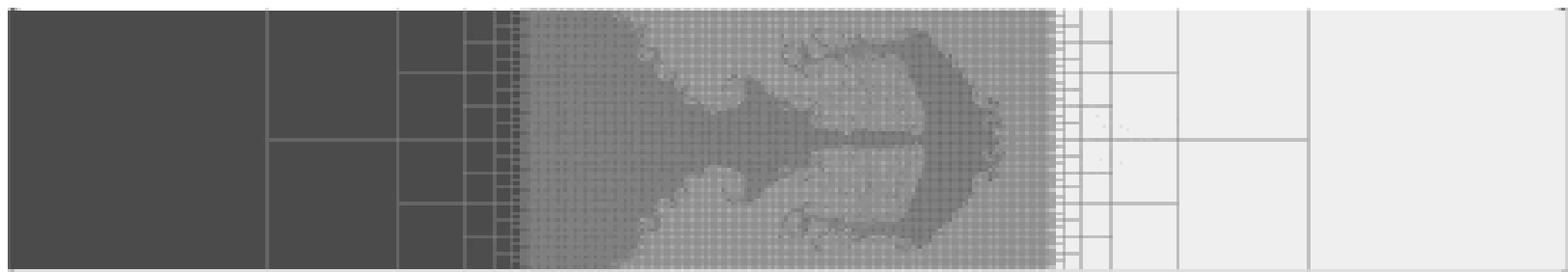}
\includegraphics[height=.625\textwidth,angle=90]{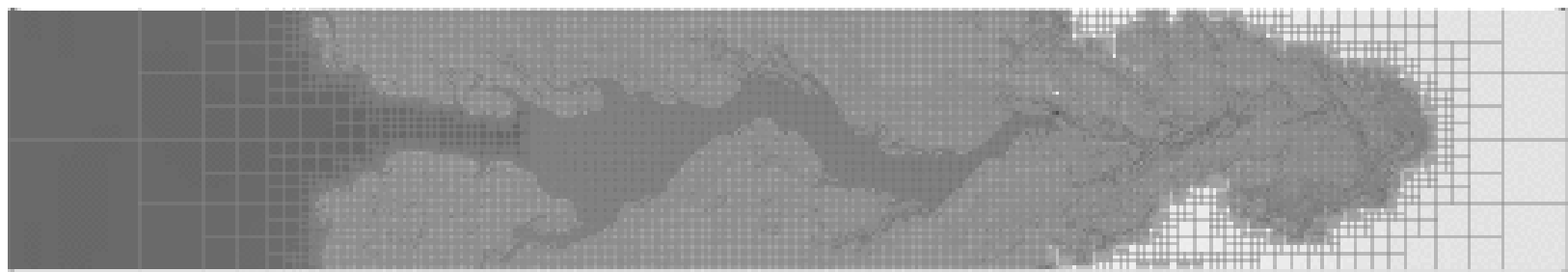}
\end{center}
\caption{Development of Rayleigh-Taylor instability at 3 epochs, from
  simulations presented in \cite{vandv}.   These are fairly high-resolution
  simulations, with a maximum of 8 levels of refinement on a top-level
  mesh with $6 \times 1$ coarsest blocks.}
\label{fig:rt}
\end{figure}

We next consider the development of the Rayleigh Taylor instability.
(Figure~\ref{fig:rt}).  This is an interface problem, but in this
set of simulations, the center region of the box is resolved to ensure
resolution of the velocity perturbations in the region near the interface.
Because this region is fully refined, many `full cost' finest
blocks are added.   This decreases the scope of improvement from TR,
as seen in Table~\ref{tab:rtresults}.

\begin{table}[hHt]
\begin{center}
\begin{tabular}{c|rlll}
time & $\nblocks$ & $R$ & $\ramr$ \\
\hline
0.0 & 33150 & 0.993 & 0.337 \\
1.8 & 33150 & 0.993 & 0.337 \\
3.6 & 60816 & 0.987 & 0.619 
\end{tabular}
\end{center}
\caption{Numerical results from simulations of a Rayleigh-Taylor instability,
         shown in Figure~\ref{fig:rt}.}
\label{tab:rtresults}
\end{table}

\section{Conclusion}

We have considered efficiency gains for time subcycling for explict or
local physics.  In these cases the work per block is roughly constant.
Further, in most cases there are many more fine blocks than coarse
blocks --- this is due to simple geometry, as a mesh that refines a
significant fraction of its domain will be strongly weighted in favour
of small blocks, which must be evolved at a small timestep.  Thus, Any
attempt to improve performance by focusing on the relatively few larger
blocks can only reduce a small fraction of the work that needs to be
done to evolve the system one timestep.  On the other hand, in studies
where only a small number of points in a large domain must be fully
resolved, there may be significant efficiency gains from TR methods.
Some cosmological hydrodynamical simulations \cite{normanextreme} are
examples of this situation.

We have not considered here accuracy; taking fewer timesteps may
increase accuracy with some solvers, although this isn't clear for
moderately time-accurate algorithms having errors of $O({\Delta t}^p)$,
$p > 1$; further, the coarsely refined regions which would benefit from
the fewer timesteps are presumably coarsely refined because the overall
solution quality is less sensitive to the error in those regions than
it is to that of the highly refined parts of the domain.
We also do not consider global or implicit solves, where the
timestepping algorithm in Fig.~\ref{fig:tmrwcurve} must be modified.
Global or implicit solves will, depending on the methods used, change
the amount of work done per block at different levels of refinement,
which can change the results given here considerably.

We have modelled only computational cost in this work.    Most of the
other costs, cf.~\S\ref{sec:analysis}, work to decrease the efficiency
gains of TR.   One unmodelled effect that could increase the gains is
the reduction of guardcell fills on large blocks.  For the oct-tree
mesh, where the number of zones per block is fixed, the reduction in
guardcell filling work is reduced in the same way as the computational
work, so that our conclusions are unchanged.   For the patch-based mesh,
the effect on the guardcell filling will be dependant on the shape of
the refined region and the algorithm used for merging patches of the
same refinement level, so that it is difficult to say anything in general.

Thus, block-structured TR significantly enhances performance of local
or explicit physics solvers only under fairly narrow circumstances.
In circumstances where TR is unlikely to produce much performance
enhancement, the added code complexity, memory overhead, and parallel
load-balancing issues may make the costs of the technique exceed its
benefits.

The authors thank B. Fryxell for useful discussions with this paper,
and K. Olson with his help with \Paramesh over the past years.  We thank
A. Calder for data from RT simulations, and F. X.  Timmes for data
from cellular detonation simulations.  We thank T. Plewa, G. Weirs,
R. Kirby, and R. Loy for suggesting this work.  Support for this work
was provided by the Scientific Discovery through Advanced Computing
(SciDAC) program of the DOE, grant number DE-FC02-01ER41176 to the
Supernova Science Center/UCSC.   LJD was supported by the Department of
Energy Computational Science Graduate Fellowship Program of the Office
of Scientific Computing and Office of Defense Programs in the Department
of Energy under contract DE-FG02-97ER25308.

The \FLASH code is freely available at http://flash.uchicago.edu/.

\bibliographystyle{plain}
\bibliography{mesh}

\end{document}